\RequirePackage[2020-02-02]{latexrelease}
\documentclass[12pt,a4paper,final]{revtex4}

\usepackage{sidecap}

\usepackage{ulem}
\usepackage{epsfig}
\usepackage{amsmath,amssymb,amsthm,mathtools}
\usepackage{graphicx}
\usepackage{bm}
\usepackage{color,soul}

\DeclareMathAlphabet{\mathpzc}{OT1}{pzc}{m}{it}

\begin{document}

\renewcommand{\textfraction}{0.00}


\newcommand{\pT}{p_\perp}
\newcommand{\vAi}{{\cal A}_{i_1\cdots i_n}}
\newcommand{\vAim}{{\cal A}_{i_1\cdots i_{n-1}}}
\newcommand{\vAbi}{\bar{\cal A}^{i_1\cdots i_n}}
\newcommand{\vAbim}{\bar{\cal A}^{i_1\cdots i_{n-1}}}
\newcommand{\htS}{\hat{S}}
\newcommand{\htR}{\hat{R}}
\newcommand{\htB}{\hat{B}}
\newcommand{\htD}{\hat{D}}
\newcommand{\htV}{\hat{V}}
\newcommand{\cT}{{\cal T}}
\newcommand{\cM}{{\cal M}}
\newcommand{\cMs}{{\cal M}^*}
\newcommand{\vk}{\vec{\mathbf{k}}}
\newcommand{\bk}{\bm{k}}
\newcommand{\kt}{\bm{k}_\perp}
\newcommand{\kp}{k_\perp}
\newcommand{\km}{k_\mathrm{max}}
\newcommand{\vl}{\vec{\mathbf{l}}}
\newcommand{\bl}{\bm{l}}
\newcommand{\bK}{\bm{K}}
\newcommand{\bb}{\bm{b}}
\newcommand{\qm}{q_\mathrm{max}}
\newcommand{\vp}{\vec{\mathbf{p}}}
\newcommand{\bp}{\bm{p}}
\newcommand{\vq}{\vec{\mathbf{q}}}
\newcommand{\bq}{\bm{q}}
\newcommand{\qt}{\bm{q}_\perp}
\newcommand{\qp}{q_\perp}
\newcommand{\bQ}{\bm{Q}}
\newcommand{\vx}{\vec{\mathbf{x}}}
\newcommand{\bx}{\bm{x}}
\newcommand{\tr}{{{\rm Tr\,}}}
\newcommand{\sNN}{s_{\mathrm{NN}}}
\newcommand{\bc}{\textcolor{blue}}

\newcommand{\beq}{\begin{equation}}
\newcommand{\eeq}[1]{\label{#1} \end{equation}}
\newcommand{\ee}{\end{equation}}
\newcommand{\bea}{\begin{eqnarray}}
\newcommand{\eea}{\end{eqnarray}}
\newcommand{\beqar}{\begin{eqnarray}}
\newcommand{\eeqar}[1]{\label{#1}\end{eqnarray}}
\newcommand{\tcr}{\textcolor{red}}

\newcommand{\half}{{\textstyle\frac{1}{2}}}
\newcommand{\ben}{\begin{enumerate}}
\newcommand{\een}{\end{enumerate}}
\newcommand{\bit}{\begin{itemize}}
\newcommand{\eit}{\end{itemize}}
\newcommand{\ec}{\end{center}}
\newcommand{\bra}[1]{\langle {#1}|}
\newcommand{\ket}[1]{|{#1}\rangle}
\newcommand{\norm}[2]{\langle{#1}|{#2}\rangle}
\newcommand{\brac}[3]{\langle{#1}|{#2}|{#3}\rangle}
\newcommand{\hilb}{{\cal H}}
\newcommand{\pleft}{\stackrel{\leftarrow}{\partial}}
\newcommand{\pright}{\stackrel{\rightarrow}{\partial}}

\newcommand{\meqbox}[2]{\eqmakebox[#1]{$\displaystyle#2$}}

\title{Probing the shape of the quark-gluon plasma droplet via event-by-event QGP tomography}

\author{Bithika Karmakar}
\affiliation{Institute of Physics Belgrade, University of Belgrade, Serbia}
\affiliation{Incubator of Scientific Excellence---Centre for
  Simulations of Superdense Fluids, University of Wroc\l{}aw,
  Poland  }

\author{Dusan Zigic}
\affiliation{Institute of Physics Belgrade, University of Belgrade, Serbia}

\author{Pasi Huovinen}
\affiliation{Incubator of Scientific Excellence---Centre for Simulations of Superdense Fluids, University of Wroc\l{}aw, Poland  }

\author{Marko Djordjevic}
\affiliation{Faculty of Biology, University of Belgrade, Serbia}

\author{Magdalena Djordjevic}
\affiliation{Institute of Physics Belgrade, University of Belgrade, Serbia}

\author{Jussi Auvinen\footnote{E-mail: jussi.a.m.auvinen@jyu.fi}}
\affiliation{University of Jyvaskyla, Finland}

\begin{abstract}
This study investigates Quark-Gluon Plasma (QGP) in heavy-ion
collisions through two avenues: high-$p_{\perp}$ frameworks and
hydrodynamic modeling. Using the T$_{\text{R}}$ENTo model, we find
that IP-Glasma mimicking $p=0$ value aligns well with high-$p_{\perp}$
data, in agreement with Bayesian analysis of the low-$p_{\perp}$
regime. While adjusting $p$ values may improve a fit to a particular
high-$p_{\perp}$ observable, it does not permit an earlier onset of
transverse expansion.
\end{abstract}

\maketitle

\section{Introduction}

A new state of matter, called Quark-Gluon Plasma (QGP), consisting of
deconfined but interacting quarks, antiquarks, and gluons, is formed
in ultra-relativistic heavy ion collisions at the Relativistic Heavy
Ion Collider (RHIC) and the Large Hadron Collider (LHC)
experiments~\cite{QGP1,QGP2,QGP3,QGP4}. These collisions create an
expanding fireball of quarks and gluons, which thermalizes to form QGP
and eventually hadronizes. Extracting useful information about the
properties of QGP formed in these collisions from the measured
particle distributions, requires comparing theoretical predictions
with the data, using phenomenological models describing the
evolution of the collision system. In particular fluid-dynamical
models have been very successful in describing a large amount of the
low-$\pT$ data.

Fluid-dynamical models do not provide the initial state of the
fluid-dynamical evolution, but it must be provided by some other
model, or parametrization. This ambiguity causes significant
uncertainty in the properties of QGP extracted from the data (see
e.g.~Ref~\cite{Song:2010mg}.) Recently, Bayesian analyses of the data
have significantly reduced this uncertainty by showing that the data
strongly favors initial entropy deposition proportional to the
geometric mean of the local nuclear thickness
function~\cite{Bernhard:2019bmu,Nijs:2020roc}. Such proportionality
corresponds to the value of the $p$ parameter of the
T$_{\text{R}}$ENTo model~\cite{Moreland:2014oya} being approximately
zero, and is consistent with the state-of-the-art
IP-Glasma~\cite{Schenke:2012wb,Schenke:2012hg} and
EKRT~\cite{Eskola:1999fc,Paatelainen:2012at,Paatelainen:2013eea}
models for initial particle production. However, these analyses are
based on low-$\pT$ particles, most of which leave the system at the
very end of the evolution, and thus reflect its properties at the late
stage.  High-$\pT$ particles formed in the primary nucleus-nucleus
collisions, on the other hand, leave the system during the first few
fm/$c$ of its evolution, and probe the initial stages of the system
more directly than low-$\pT$ particles. Therefore, it is important to
study whether the high-$\pT$ observables, $R_{AA}$ and flow harmonics
$v_2$, $v_3$, and $v_4$, favor the same initial shape of the fireball
as the low-$\pT$ particles.

Furthermore, in our previous works~\cite{Stojku:2020wkh,Stojku:2023mze}
we found that the high-$\pT$ observables can be reproduced
only if the transverse expansion is very weak (or nonexistent) until
$\tau\approx 1$ fm, but the Bayesian analyses using the T$_{\text{R}}$ENTo
model usually assume immediate transverse expansion after the primary
collisions in terms of free-streaming~\cite{Bernhard:2019bmu,Nijs:2020roc}.
This brings into question whether delaying transverse expansion until
$\tau_0 = 1$ fm will change the shape favored by the data. As well,
the requirement of late onset of transverse expansion was based on
conventionally shaped initial states corresponding to the $p\approx 0$
value. Thus, it is possible that if the initial state was way more
anisotropic than commonly assumed, transverse expansion could begin
earlier, and high-$\pT$ anisotropies would still be reproduced.

In this work, we address these two issues: We modulate the initial
shape of the system by altering the T$_{\text{R}}$ENTo parameter $p$
while adjusting the other parameters to reproduce a subset of the
low-$\pT$ data. We do not allow transverse expansion before the onset
of fluid-dynamical evolution at $\tau_0 = 1$ fm, calculate the
high-$\pT$ observables for each value of $p$, and compare with the
data to see which value, and thus shape, the data favor. We also
systematically decrease the $p$-value down to $p=-3$ to select more
anisotropic initial distribution and carry out calculations using
$\tau_0 = 0.6$ fm and $\tau_0 = 0.2$ fm to see if highly anisotropic
initial state would lead to large enough high-$\pT$ anisotropies even
if the early transverse expansion diluted a large part of the spatial
anisotropy by the time the jets become sensitive to it.

\section{Methods}
\subsection{Bulk medium evolution}

To generate the event-by-event fluctuating initial entropy profiles,
we use the phenomenological model
T$_{\text{R}}$ENTo~\cite{Moreland:2014oya}, which produces the initial
profiles without assuming any specific mechanism for entropy
production. Instead, the starting point is the so-called thickness
function of the colliding nuclei $A$ and $B$
\bea
  \tilde T_{A,B}(x,y)=\int dz \rho_{A,B}^{\text{part}}(x,y,z),
\eea
where $\rho_{A,B}^{\text{part}}$ is the density of the nuclear matter
that participates in the inelastic collisions. In event-by-event case
it is calculated by sampling the Woods-Saxon distribution to create
the positions of nucleons, which are described as Gaussian
distributions. For this process, we use the parameters from
Ref.~\cite{Karmakar:2023bxq}: width $w = 0.5$ fm, minimum
nucleon-nucleon distance $d = 0.5$ fm, inelastic nucleon-nucleon
cross section $\sigma_{NN} = 7$ fm$^2$ and fluctuation parameter
$k=1.19$. The initial entropy density is related to the thickness
functions as
\bea
  s(x,y) \propto \tilde T_R(p; \tilde T_A, \tilde T_B)
         \equiv \left( \frac{\tilde T_A^p+\tilde T_B^p}{2} \right)^{1/p},
  \label{trento_f}
\eea
where $p$ is a continuous parameter which can have values from
$-\infty$ to $\infty$. As known, T$_{\text{R}}$ENTo can be used to
mimic and interpolate between various initial state models; the values
of $p$ $-2/3, 0$ and $1$ corresponding to KLN, EKRT/IP-Glasma, and
wounded nucleon models, respectively. In this study, however, we
consider $p$ a free parameter that allows us to test different shapes
of the initial state. In particular, we use $p \in \{1/3, 0, -1/3,
-2/3, -1, -3\}$ to see which value the high-$\pT$ data favors and
whether higher initial anisotropy would allow the earlier onset of
transverse expansion.

We do not allow any pre-equilibrium evolution (free-streaming or
otherwise). In the first part of the study, we use $\tau_0 = 1$ fm as
the initial time for fluid-dynamical evolution with zero transverse
velocity since the reproduction of the high-$\pT$ observables does not
allow strong transverse expansion earlier~\cite{Stojku:2020wkh}. In
the second part of the study, we use $\tau_0 = 0.6$ fm and
$\tau_0 = 0.2$ fm to test whether large anisotropy would allow the
earlier onset of transverse expansion.

We evolve the event-by-event fluctuating initial distributions using
the version of VISHNew~\cite{Song:2007ux,Song:2008si} used in
Refs.~\cite{Bernhard:2016tnd,Bernhard:2018hnz,Bernhard:2019bmu}\footnote{Code available at~\cite{Jonahs_github}}.
This code solves the dissipative fluid-dynamical equations in
2+1-dimensions, i.e., assuming boost invariance. Shear stress and bulk
pressure are taken as dynamical variables and evolved according to the
Israel-Stewart type equations~\cite{Israel:1979wp}. We use an Equation
of State (EoS)~\cite{Bernhard:2016tnd} that combines the lattice
QCD-based EoS of the HotQCD collaboration~\cite{HotQCD:2014kol} at
large temperatures to a hadron resonance gas EoS at low temperatures.
The fluid is particlized on an isothermal hypersurface of
$T_{\mathrm{sw}} = 151$ MeV using the Cooper-Frye
prescription~\cite{Cooper:1974mv}, and the particles fed to the UrQMD
hadron cascade~\cite{Bass:1998ca,Bleicher:1999xi}, which describes the
evolution of the hadronic stage of the system.

The temperature dependence of the specific bulk viscosity is
parameterized as a Cauchy distribution~\cite{Bernhard:2018hnz} with the
maximum value $(\zeta/s)_{\text{max}}$ = 0.03, and the other
parameters characterizing the bulk viscosity being the same as in
Ref.~\cite{Karmakar:2023bxq}. Instead of allowing
temperature-dependent specific shear viscosity, we use constant
$\eta/s$, since in our earlier study~\cite{Karmakar:2023bxq} we found
that the high-$\pT$ observables could be equally well reproduced using
constant $\eta/s$ as temperature dependent $\eta/s$.

\begin{table}[t]
  \begin{center}
    \begin{tabular}{|c|c c c c c c||c c||c c|}\hline
      & \multicolumn{6}{c||}{ $\tau_0 = 1$ fm } &
        \multicolumn{2}{c||}{\,$\tau_0 = 0.6$ fm\, } &
        \multicolumn{2}{c|}{\,$\tau_0 = 0.2$ fm\, } \\ \hline
      $p$      & 1/3  & 0    & -1/3 & -2/3 & -1   & -3   & 0  & -3
               & 0    & -3  \\
        \hline
      $\eta/s$ & \,0.03\, & 0.15\, & 0.18\, & 0.20\, & 0.25\, & 0.29\,
               & \,0.17 & 0.33     & \,0.20 & 0.60\\
        \hline  
    \end{tabular}
    \caption{$\eta/s$ values used for each value of $p$ and
             initial time $\tau_0$}
      \label{eta_params}
  \end{center}
\end{table}

We keep all the other parameters fixed but adjust the
T$_{\text{R}}$ENTo normalization parameter and $\eta/s$ for each value
of the $p$ parameter to reproduce the charged particle elliptic flow
$v_2\{4\}$, and proton, pion and kaon multiplicities in each
centrality class. The required values of $\eta/s$ are shown in
Table~\ref{eta_params}.

We generate $10^4$ events for each T$_{\text{R}}$ENTo $p$ value. These
events are then sorted into centrality bins according to the number of
participants in the collision. We analyze the events in four
centrality classes, 10-20\%, 20-30\%, 30-40\%, and 40-50\%, where the
low-$p_\perp$ observables are calculated for each event and averaged
over the events in the class. Subsequently, the temperature profiles
in each centrality class are used as input to the generalized DREENA-A
to calculate the high-$p_\perp$ observables in each class.

\subsection{Generalized DREENA-A framework}

In our study, the high-$p_\perp$ energy loss has been computed using
the DREENA-A (Dynamical Radiative and Elastic ENergy loss Approach)
framework where $`$A' stands for Adaptive, {\it{i.e.,}} arbitrary
temperature profile can be included as input in this framework. This
has been further optimized to incorporate event-by-event fluctuations
in the temperature profiles. The framework is based on generalized
hard thermal loop perturbation theory~\cite{Kapusta:1989tk} in which
the infrared divergences are naturally
regulated~\cite{Djordjevic:2006tw,Blagojevic:2018nve}. It has several
important features that make it a reliable tool for QGP tomography: i)
The system is considered as finite size dynamical medium {\it{i.e.,}}
it consists of moving partons, ii) Both the
radiative~\cite{Djordjevic:2009cr,Djordjevic:2008iz} and
collisional~\cite{Djordjevic:2006tw} energy loss have been computed
within the same framework, iii) It is generalized to include the
running coupling~\cite{Djordjevic:2013xoa}, finite magnetic
mass~\cite{Djordjevic:2011dd}, and beyond soft-gluon
approximation~\cite{Blagojevic:2018nve}. iv) It does not have fitting
parameters in the energy loss, {\it{i.e.,}} all the parameters are set
to standard literature values~\cite{Stojku:2020wkh,Stojku:2021yow},
allowing systematic comparison of data and the predictions from the
simulation obtained using the same formalism and parameter set. We
recently also extended the formalism to include finite orders in
opacity~\cite{Stojku2023} but found out that including these
effects has a minor impact on high-$p_\perp$ predictions. Thus, a
computationally more efficient version with one scattering center is
used in this study.

To generate the medium-modified distribution of high-$p_\perp$
particles, we use the following pQCD convolution formula:
\bea
\frac{E_f d^3\sigma_q(H_Q)}{dp^3_f}=\frac{E_id^3\sigma(Q)}{dp^3_i}\otimes P(E_i \rightarrow E_f) \otimes D(Q\rightarrow H_Q),
\eea
where $i$ and $f$ denote the initial parton ($Q$) and the final hadron
($H_Q$) respectively. $\frac{E_id^3\sigma(Q)}{dp^3_i}$ represents
initial parton spectrum that is calculated in accordance with
Ref.~\cite{Kang:2012kc, Sharma:2009hn, Cacciari:2012ny}. As discussed
above, the energy loss probability $P(E_i \rightarrow E_f)$ has been
computed within finite temperature field theory. $D(Q\rightarrow H_Q)$
represents the fragmentation function, where DSS, BCFY, and KLP
fragmentation functions have been used for light hadrons, D, and B
mesons, respectively. We consider the gluon mass
$m_g=\mu_E/\sqrt{2}$~\cite{Djordjevic:2003be} where $\mu_E$ is the
temperature-dependent Debye mass that is calculated following the
procedure in Ref.~\cite{Peshier:2006ah}. Further, we assume the mass
of the light quark, charm, and bottom quark to be $M=\mu_E/6$, 1.2
GeV, and 4.75 GeV, respectively. The magnetic to electric mass ratio
is $\mu_M/\mu_E=0.6$~\cite{Borsanyi:2015yka}. Also, we assume
$\Lambda_{QCD} = 0.2$ GeV and effective number of light quark flavors
$n_f=3$ in our study.

\section{Results}

\begin{center}
\begin{figure}[tbh!]
 \begin{center}
 \includegraphics[scale=0.4]{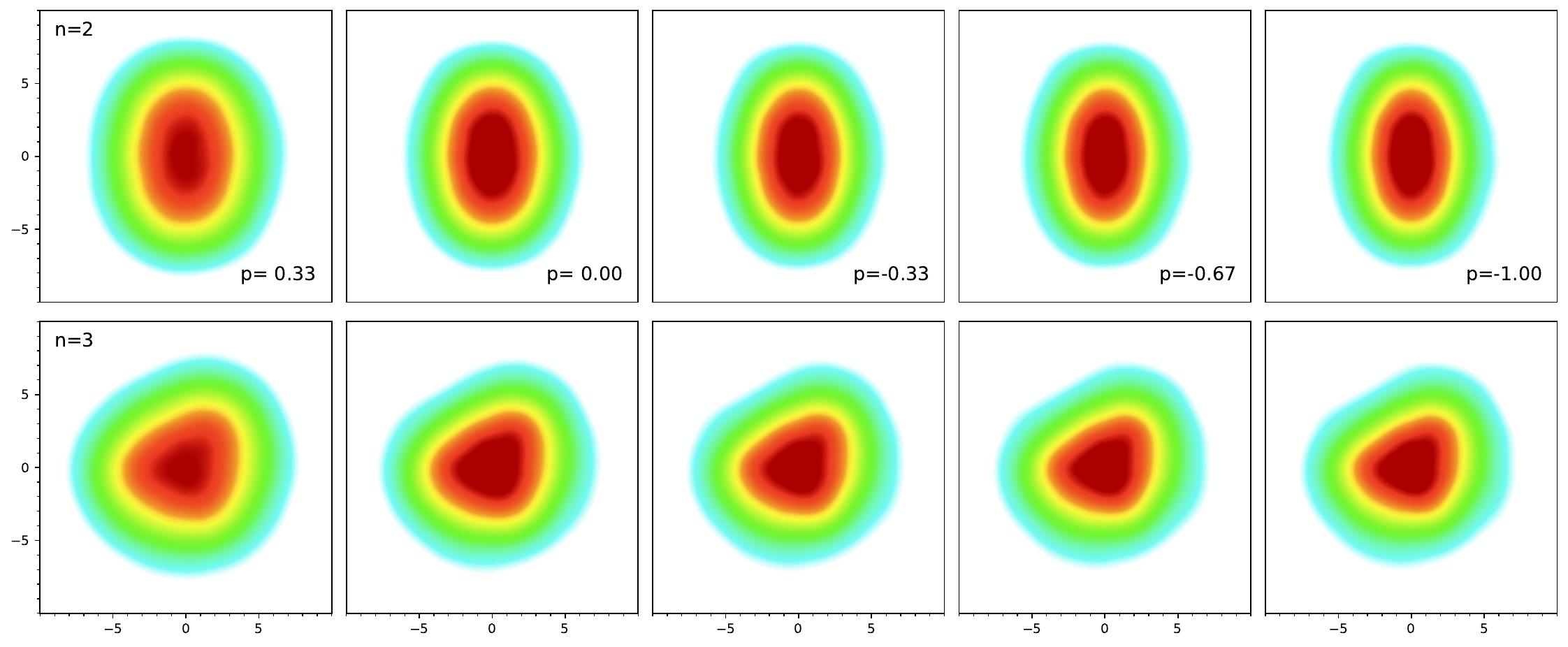}
 \caption{The averaged initial ($\tau=1$ fm) temperature distributions for
   $p\in\{1/3,0,-1/3,-2/3,-1\}$ at 20-30\% centrality. When averaging,
   the participant angles of each event were aligned. The
   distributions in the upper row were aligned with respect to the
   $\Psi_{2,2}$ angle, and in the lower row w.r.t.~$\Psi_{2,3}$,
   making the average ellipticity and triangularity visible.}
  \label{temp_ini}
 \end{center}
\end{figure}
\end{center}

\begin{figure}
  \includegraphics[scale=0.4]{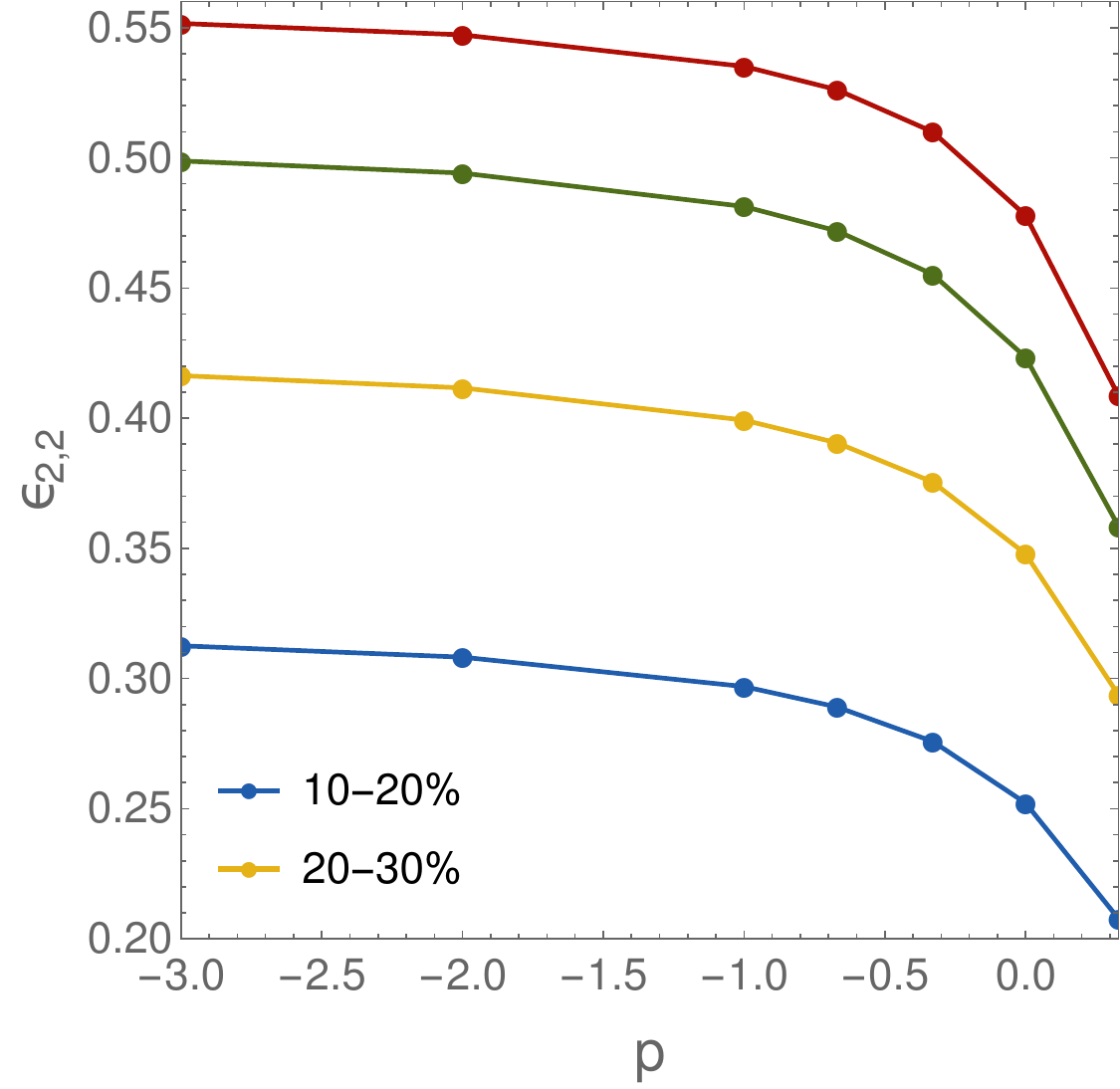}
  \includegraphics[scale=0.4]{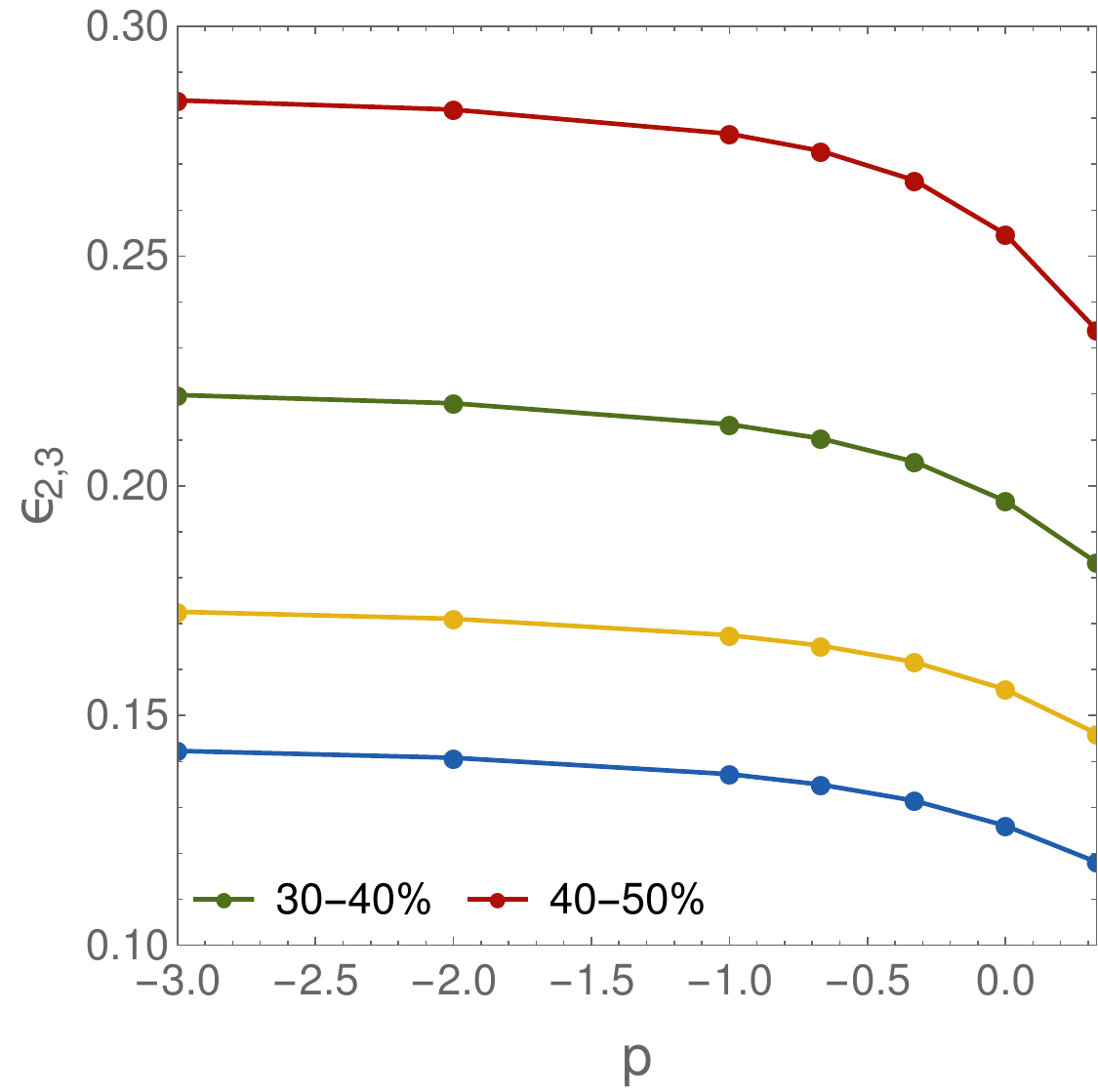}
  \caption{The anisotropy parameters $\epsilon_{2,2}$ and
    $\epsilon_{2,3}$ as function of $p$ at various centralities.}
  \label{anisotropies}
\end{figure}

We characterise the initial state anisotropies with the coefficients $\epsilon_{m,n}$ ~\cite{Niemi:2012aj}
\begin{equation}
 \epsilon_{m,n}=-\frac{\int dx dy\, r^m\cos \left[ n(\phi-\Psi_{m,n}) \right] e(x,y)}{\int dx dy\, r^m e(x,y)}
\end{equation}
where $(r, \phi)$ are the radial coordinates in the transverse plane, $e(x,y)$ is the energy density and $\Psi_{m,n}$ is the event plane angle:
\begin{equation}
 \Psi_{m,n}=\frac{1}{n}\arctan \frac{\int dx dy\, r^m \sin (n\phi) e(x,y)}{\int dx dy\, r^m \cos (n\phi) e(x,y)} + \pi/n.
\end{equation}

To illustrate how the shape of the initial state depends on the
parameter $p$, we show in Fig.~\ref{temp_ini} the averaged initial
temperature distributions at 20-30\% centrality for various values of
$p$, and the anisotropy parameters $\epsilon_{2,2}$ and
$\epsilon_{2,3}$ as function of $p$ at various centralities in
Fig.~\ref{anisotropies}.  When averaging, the individual events were
rotated to align their participant angles $\Psi_{2,2}$ (upper row of
Fig.~\ref{temp_ini}) or $\Psi_{2,3}$ (lower row of
Fig.~\ref{temp_ini}) to make the average
ellipticity (upper row) or triangularity (lower row) visible. As seen,
the system becomes more elliptically elongated and more triangular
when $p$ is reduced.

\begin{figure}[tbh!]
 \includegraphics[scale=0.4]{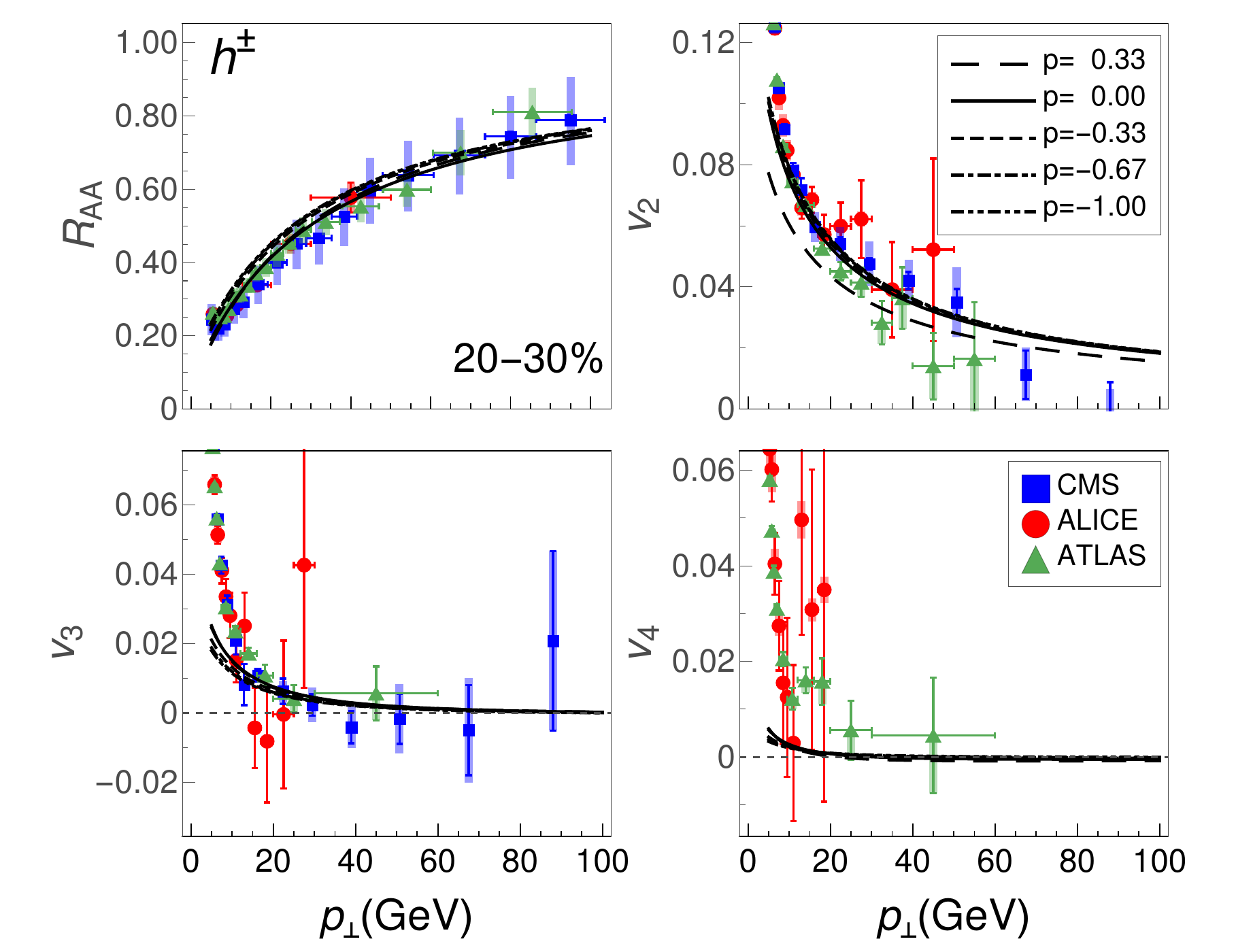}
 \caption{The charged hadron $R_{AA}$ and high-$p_\perp$ flow
   harmonics $v_2$, $v_3$ and $v_4$ as function of transverse momentum
   in Pb+Pb collisions at $\sqrt{s_{\text{NN}}}=5.02$ TeV for
   different initializations at 20-30\% centrality. CMS (blue
   squares)~\cite{CMS:2017xgk, CMS:2016xef}, ALICE (red
   circles)~\cite{ALICE:2018vuu, ALICE:2018rtz}, and ATLAS (green
   triangles)~\cite{ATLAS:2018ezv, ATLAS:2017rmz} experimental data
   are also shown for comparison.}
  \label{PbPb_ch_2030}
\end{figure}
\begin{figure}[b!]
 \includegraphics[scale=0.4]{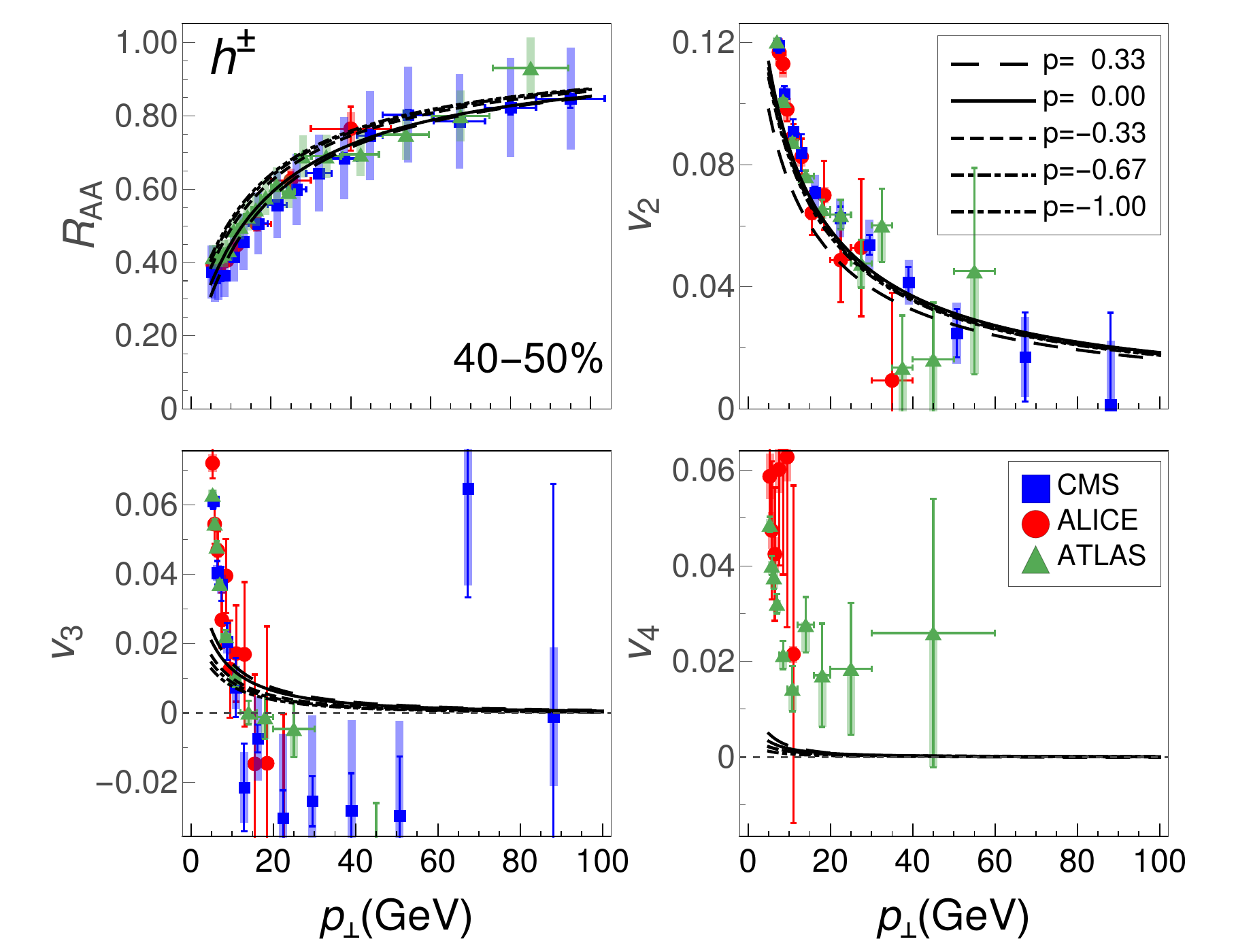}
 \caption{The same as Fig.~\ref{PbPb_ch_2030}, but at 40-50\% centrality.}
  \label{PbPb_ch_4050}
\vspace*{-20mm}
\end{figure}

At low $\pT$, the $v_2$ and $v_3$ are strongly correlated with their
corresponding spatial anisotropies~\cite{Niemi:2012aj}, but at high
$\pT$ their behavior is more complicated. This can be seen in
Figs.~\ref{PbPb_ch_2030} and~\ref{PbPb_ch_4050}, where we show the
calculated $R_{AA}$ and high-$\pT$ $v_2$, $v_3$ and $v_4$ at 20-30\%
and 40-50\% centrality, respectively. These results are for Pb+Pb
collisions at $\sqrt{s_{\text{NN}}}=5.02$~TeV, and compared with the
data. When $p$ is non-negative, high-$\pT$ $v_2$ increases with
decreasing $p$, i.e., with increasing $\epsilon_{2,2}$, but once $p$
becomes negative $v_2$ hardly changes even if $\epsilon_{2,2}$ keeps
increasing. On the other hand, high-$\pT$ $v_3$ is
\emph{anticorrelated} with $\epsilon_{2,3}$ and tends to decrease when
$\epsilon_{2,3}$ increases (i.e., when $p$ decreases), although when
$p$ is non-negative, $v_3$ hardly changes. This is surprising, but one
has to keep in mind that the high-$\pT$ anisotropies are not as sensitive
to the initial shape of the system as they are to the time-average of the
shape and the temperature of the system. Thus, at high $\pT$, we should not expect
$v_2$ nor $v_3$ to be directly related to the eccentricities $\epsilon_{2,2}$ and $\epsilon_{2,3}$.  Furthermore, as we have noted
earlier~\cite{Stojku:2021yow}, the high-$\pT$ observable which
reflects the shape of the system best is not $v_2$, but the ratio
$v_2/(1-R_{AA})$.

The $R_{AA}$ hardly depends on the shape of the system, but as we have
discussed previously~\cite{Stojku:2021yow}, the initial temperature
where the energy loss sets in dominates $R_{AA}$. The $p$ value does
not directly affect the initial temperature, but since we require our
calculations to reproduce the observed multiplicity and $v_2\{4\}$, we
have to adjust the initial temperature accordingly. As seen in
Fig.~\ref{temp_ini}, the smaller the $p$, the smaller the system.
Reaching the same final multiplicity necessitates a larger initial
entropy density and, thus, higher temperature. On the other hand, the
high initial eccentricity necessitates higher shear viscosity to dampen
the low-$\pT$ $v_2$. Higher viscosity leads to higher entropy production
during the evolution, and to obtain the same final multiplicity, the
initial entropy density and, thus, temperature should be smaller. These
two opposing trends do not quite cancel each other but lead to the
overall tendency of initial temperature increasing with increasing $p$. 
The calculated $R_{AA}$, shown in the upper left
panels of Figs.~\ref{PbPb_ch_2030} and~\ref{PbPb_ch_4050}, generally
follows the behavior of initial temperature: the larger the $p$ the
lower the $R_{AA}$. Based on these observations and comparison with the
data\footnote{Because of the large error bars and low statistics we do
not pay attention to the $v_3$ data above $\pT \approx 10$ GeV.}, we
can say that the observed $R_{AA}$ and $v_3$ favors values $p \gtrsim
0$, whereas high-$\pT$ $v_2$ favors values $p \lesssim 0$. Thus, we
conclude that the high-$\pT$ data favors the value $p \approx 0$ in
agreement with the low-$\pT$ data~\cite{Bernhard:2019bmu}.

In contrast to the previous observables, the high-$p_\perp$ $v_4$ is
almost insensitive to the $p$ value (lower right panels of
Figs.~\ref{PbPb_ch_2030} and~\ref{PbPb_ch_4050}). Furthermore, our
generalized DREENA-A framework consistently underpredicts the
experimental data for charged hadron $v_4$. This discrepancy has been
previously highlighted in our earlier works~\cite{Zigic:2022xks,
  Karmakar:2023bxq}, where variations in initial conditions, including
the presence or absence of initial transverse free streaming, and
different shear viscosity-to-entropy density ($\eta/s$)
parametrizations led to similarly low predictions for high-$p_\perp$
$v_4$. We have called this divergence between theory and experiment
the "high-$p_\perp$ $v_4$ puzzle"~\cite{Zigic:2022xks}.

Our comprehensive analyses demonstrate that attempts to resolve this
puzzle through modifying parameters in fluid dynamical
simulations---including eliminating free streaming, delaying
thermalization time, employing diverse $\eta/s$ parametrizations, or
varying the $p$ values for initial conditions---have yielded negligible
impact on high-$p_\perp$ $v_4$. If upcoming high-luminosity
experimental data corroborate these observations, resolving the $v_4$
puzzle might necessitate alternative approaches to initial state
and/or energy loss models.

\begin{center}
\begin{figure}[t]
 \begin{center}
 \includegraphics[scale=0.3]{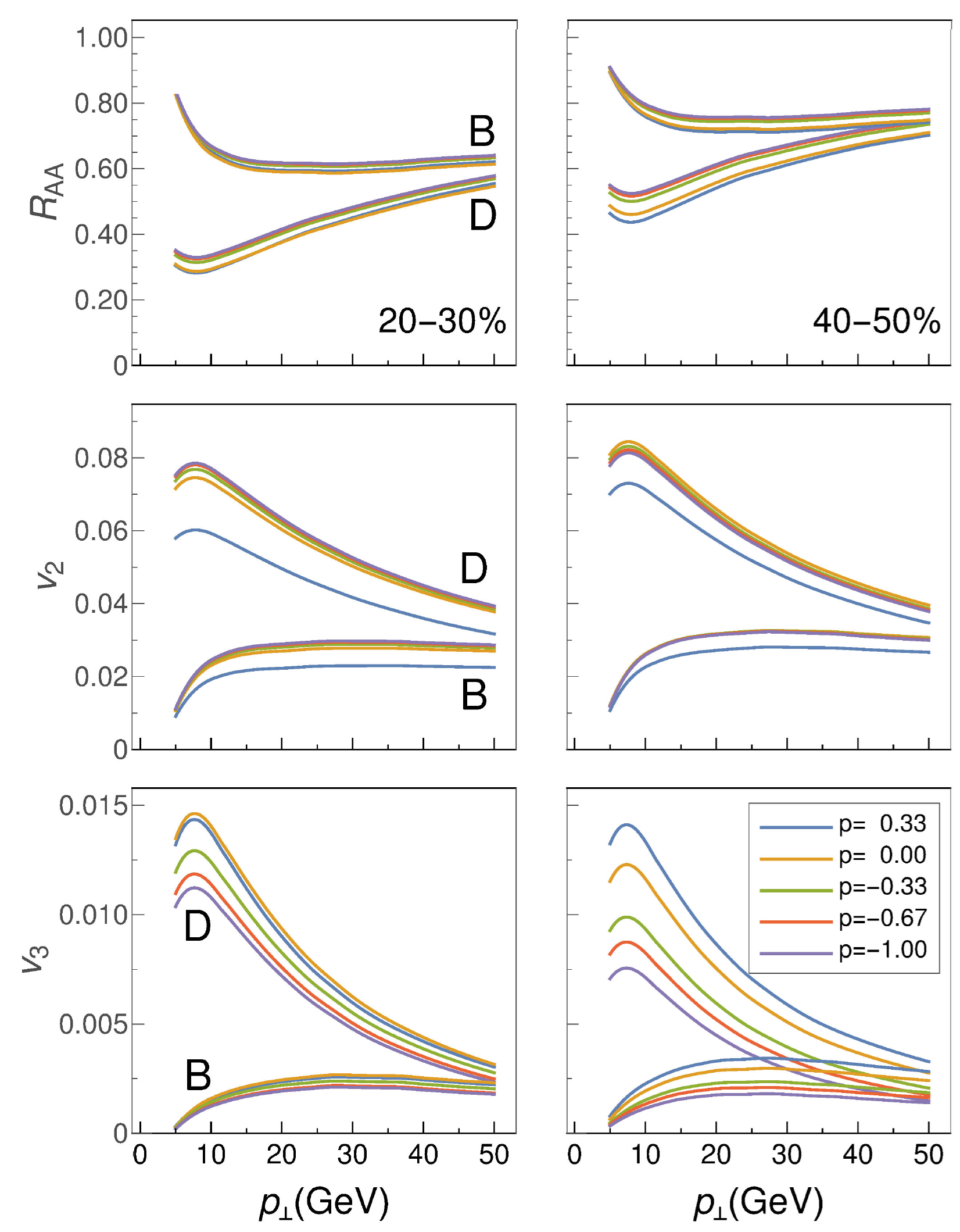}
 \caption{The D and B meson $R_{AA}$ (first row) and high-$p_\perp$
   flow harmonics $v_2$ (second row) and $v_3$ (third row) as function
   of transverse momentum in Pb+Pb collisions at
   $\sqrt{s_{\text{NN}}}=5.02$ TeV for different initializations. The
   left column depicts collisions at 20-30\% centrality and the right
   column at 50-60\% centrality.}
  \label{PbPb_DB_all}
 \end{center}
\end{figure}
\end{center}

As known, the heavy quark energy loss is more sensitive to the details
of the evolution of the fireball than the light quark energy
loss. Therefore, we have also evaluated the $R_{AA}$ and high-$\pT$
anisotropies of D and B mesons shown in Fig.~\ref{PbPb_DB_all}. The
overall behavior is the same as for charged hadrons, but in particular
$v_3$ becomes more sensitive to the initial anisotropy, and is able to
differentiate between all values of $p$ we tested. This enhanced
sensitivity to details of the evolution hints at the possibility that
once the high-luminosity data becomes available from sPHENIX and Run 3
at the LHC, the study of D and B meson anisotropies may reveal
structures so far hidden, and which the T$_{\text{R}}$ENTo model
cannot describe.

\begin{center}
\begin{figure}[t]
 \begin{center}
 \includegraphics[scale=0.3]{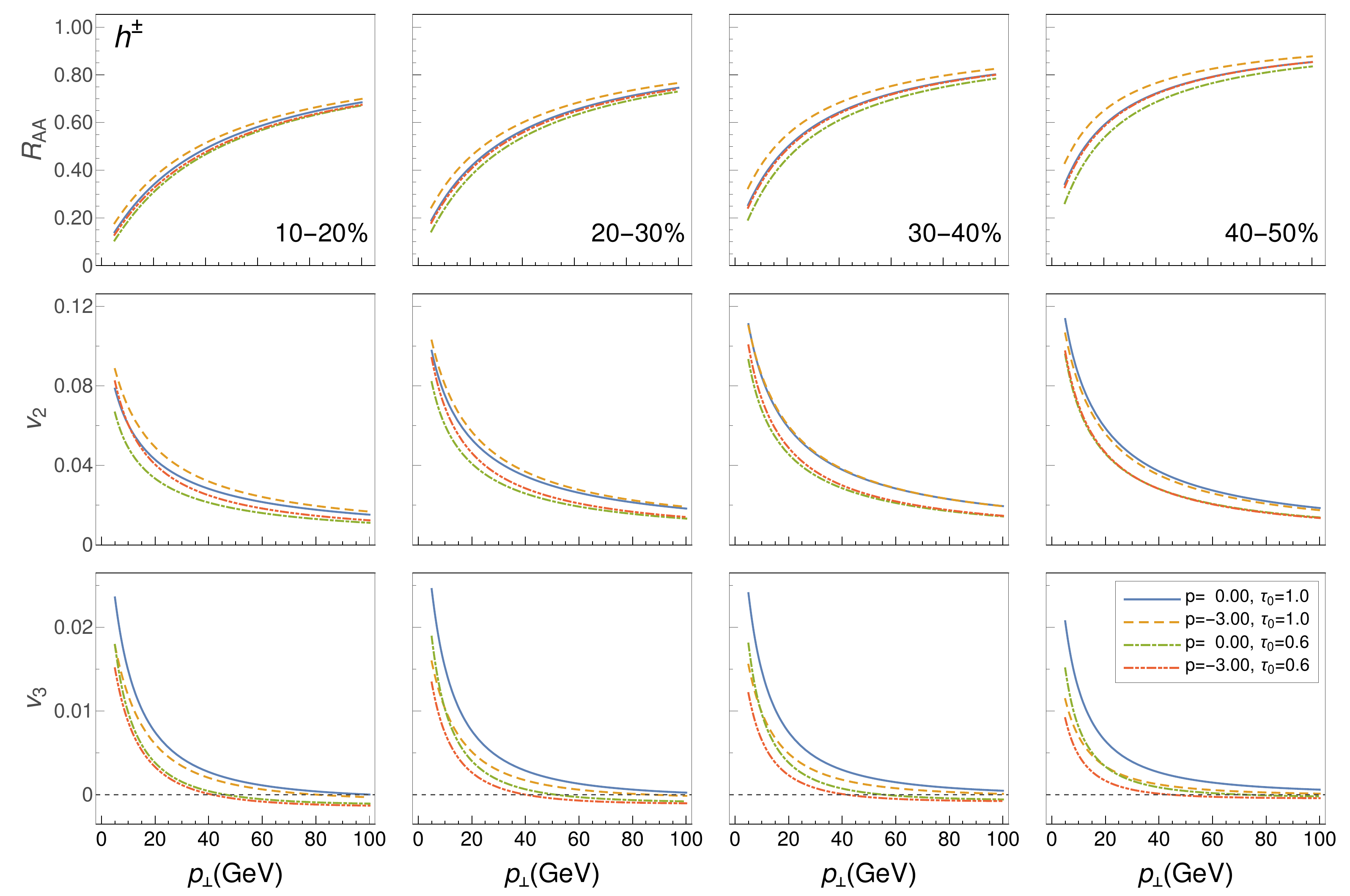}
 \caption{Calculated charged hadron $R_{AA}$ (first row),
   high-$p_\perp$ $v_2$ (second row) and $v_3$ (third row) as function
   of transverse momentum for $p = 0$ and $-3$ and $\tau_0 = 1$ and
   0.6 fm at 10-20\%, 20-30\%, 30-40\% and 40-50\% centrality classes.}
  \label{p_tau_ch}
 \end{center}
\end{figure}
\end{center}

As discussed in the Introduction, we had observed in our earlier works
that the reproduction of high-$\pT$ $v_2$ is possible only if the
transverse expansion of the fireball is delayed to $\tau\approx 1$
fm~\cite{Stojku:2020wkh}. To study whether it would still be possible
to start transverse expansion earlier and fit the $v_2$ data if the
initial anisotropy was way larger than usually assumed, we cranked the
$p$ parameter of the T$_{\text{R}}$ENTo model down to -3, and
readjusted viscosity and normalization to reproduce the low-$\pT$
data. The resulted high-$\pT$ observables, $R_{AA}$, $v_2$ and $v_3$
of charged hadrons, are shown in Fig.~\ref{p_tau_ch}. A comparison of the
solid blue ($p=0$ and $\tau_0=1$ fm) and dashed orange ($p=-3$ and
$\tau_0=1$ fm) curves reveals behavior similar to previously discussed:
Smaller $p$ reduces temperature and increases both elliptic and
triangular anisotropies. Consequently $R_{AA}$ and $v_2$ increase
whereas $v_3$ decrease. Interestingly, the increase in $R_{AA}$ might
allow an earlier start of energy loss without reducing $R_{AA}$ too
much. The red curve ($p=-3$ and $\tau_0=0.6$ fm) reveals this to be
the case. $R_{AA}$ is almost identical to $R_{AA}$ obtained using
$p=0$ and $\tau_0=1$ fm, but the increase in $v_2$ is insufficient to
compensate for the reduction due to the earlier onset of transverse
expansion. Furthermore, both small $p$ and early onset of expansion
reduce $v_3$, leading to smaller $v_3$ than experimentally
observed. To be sure of our conclusions, we repeated the calculations
using $p = -1$ and $-2$, and obtained results which were between the
ones obtained using $p=0$ and $p=-3$. One could ask what would happen
if we cranked $p$ down even further, say, down to minus eleven, but as
shown in Fig.~\ref{anisotropies}, once $p \lesssim -2$, the
anisotropies change very little. Therefore, we dare to conclude that
if there is an initial state which is anisotropic enough to allow
both early onset of transverse expansion and reproduction of high-$\pT$
anisotropies, it cannot be modeled using T$_{\text{R}}$ENTo.

As mentioned in the Introduction and Section IIA, we repeated this
study using an even earlier start of transverse expansion and jet
energy loss at $\tau=0.2$ (results not shown), and it was even clearer
that the increased anisotropy was insufficient to compensate for the
reduction of $R_{AA}$ and high-$\pT$ $v_2$ due to early onset of
expansion and energy loss~\footnote{We remind the reader that
in~\cite{Stojku:2020wkh}, we studied the early expansion using both
free streaming and fluid dynamics, and the results were the same:
Early transverse expansion strongly reduces high-$\pT$ $v_2$
irrespectively whether the expansion is fluid dynamical or not.}.

\section{Summary}

In the fluid-dynamical models of heavy-ion collisions, the initial
state is conventionally constrained by comparison to low-$\pT$
data. In this study, we have provided a complementary approach by using
high-$\pT$ data, i.e., QGP tomography, to constrain the initial
state. In particular, we carried out two distinct investigations:

i) We modulated the shape of the initial state by varying the
parameter $p$ of the T$_{\text{R}}$ENTo model. We found out that while
$R_{AA}$ and $v_3$ preferred values $p \gtrsim 0$, high-$\pT$ $v_2$ preferred
$p \lesssim 0$. Thus, the best overall fit was obtained
with $p \approx 0$, consistently with the Bayesian analysis of low-$\pT$ data.

ii) We tested if larger elliptic anisotropy of the initial state,
i.e., $p \ll 0$ would allow an earlier onset of transverse, in our case
fluid-dynamical, expansion. Our results show that, while lower $p$
values enhance high-$p_\perp$ $R_{AA}$ and $v_2$, the enhancement is
insufficient for facilitating an earlier onset of transverse expansion
in heavy-ion collisions.

We conclude by noting that by providing a complementary approach to
constrain the early stages of heavy-ion collisions, this research
demonstrates the value of employing jet tomography, in which both low-
and high-$p_\perp$ theoretical and experimental data are collectively
utilized to refine our understanding of QGP properties.

\underline{Acknowledgements:} This work is supported by the European Research Council, grant ERC-2016-COG: 725741, and by the Ministry of Science and Technological Development of the Republic of Serbia. PH and BK were also supported by the program Excellence Initiative–Research University of the University of Wroclaw of the Ministry of Education and Science. JA acknowledges the financial support from the Academy of Finland Project No. 330448. JA's research was also funded as a part of the Center of Excellence in Quark Matter of the Academy of Finland (Project No. 346325). This research is part of the European Research Council Project No. ERC-2018-ADG-835105 YoctoLHC.

\end{document}